\documentclass[journal]{IEEEtran}

\usepackage[utf8]{inputenc}
\usepackage{color}
\usepackage{array}
\usepackage{verbatim}
\usepackage{float}
\usepackage{amsmath}
\usepackage{amsthm}
\usepackage{amssymb}
\usepackage{graphicx}
\usepackage{longtable}
\usepackage{multirow}
\usepackage[unicode=true,
bookmarks=false,
breaklinks=false,pdfborder={0 0 1},colorlinks=false]
{hyperref}
\hypersetup{
	colorlinks,bookmarksopen,bookmarksnumbered,citecolor=blue,urlcolor=blue}
\usepackage{cite}

\floatstyle{ruled}
\newfloat{algorithm}{tbp}{loa}
\providecommand{\algorithmname}{Algorithm}
\floatname{algorithm}{\protect\algorithmname}

\makeatletter
\let\oldforeign@language\foreign@language
\DeclareRobustCommand{\foreign@language}[1]{%
	\lowercase{\oldforeign@language{#1}}}

\let\oldforeign@language\foreign@language
\DeclareRobustCommand{\foreign@language}[1]{%
	\lowercase{\oldforeign@language{#1}}}

\ifCLASSINFOpdf
\else
\fi

\newcommand{\MYfooter}{\smash{
		\hfil\parbox[t][\height][t]{\textwidth}{\centering
			\thepage}\hfil\hbox{}}}

\makeatletter

\def\ps@IEEEtitlepagestyle{%
	\def\@evenhead{\scriptsize\thepage \hfil \leftmark\mbox{}}%
	\def\@oddfoot{\parbox[t][\height][l]{\textwidth}{
			\vspace{-20pt}{\rule{\textwidth}{0.4pt}}\\ \footnotesize
			 This paper has been accepted for publication in the 2020 IEEE 24th International Conference on System Theory, Control and Computing (ICSTCC), Sinaia, Romania. The paper has already been presented by Trevor P. Drayton.\\
			\noindent\makebox[\linewidth]
		}\hfil\hbox{}}%
	\def\@evenfoot{\MYfooter}}
\newtheorem{defn}{Definition}

\newtheorem{thm}{Theorem}
\newtheorem{rem}{Remark}

\newtheorem{assum}{Assumption}

\begin{document}
	\bstctlcite{IEEEexample:BSTcontrol}

	\title{Fast Adaptation Nonlinear Observer for SLAM}
	
\author{Trevor P. Drayton, Abdul A. Jaiyeola, Nazmul Hoque, Mikhayla Maurer, and~Hashim~A.~Hashim\\
	Software Engineering\\
	Department of Engineering and Applied Science\\
	Thompson Rivers University,	Kamloops, British Columbia, Canada, V2C-0C8\\
	draytont10@mytru.ca, jaiyeolaa17@mytru.ca, hoquen18@mytru.ca, maurerm18@mytru.ca and hhashim@tru.ca
		\thanks{This work was supported in part by Thompson Rivers University Internal
			research fund, RGS-2020/21 IRF, \# 102315.}
		
	}
	
	
	\maketitle
	
	\begin{abstract}
The process of simultaneously mapping the environment in three dimensional
(3D) space and localizing a moving vehicle's pose (orientation and
position) is termed Simultaneous Localization and Mapping (SLAM).
SLAM is a core task in robotics applications. In the SLAM problem,
each of the vehicle's pose and the environment are assumed to be completely
unknown. This paper takes the conventional SLAM design as a basis
and proposes a novel approach that ensures fast adaptation of the
nonlinear observer for SLAM. Due to the fact that the true SLAM problem
is nonlinear and is modeled on the Lie group of $\mathbb{SLAM}_{n}\left(3\right)$,
the proposed observer for SLAM is nonlinear and modeled on $\mathbb{SLAM}_{n}\left(3\right)$.
The proposed observer compensates for unknown bias attached to velocity
measurements. The results of the simulation illustrate the robustness
of the proposed approach.
	\end{abstract}
	

	\IEEEpeerreviewmaketitle{}

	\section{Introduction}
	
	\IEEEPARstart{S}{imultaneous} Localization and Mapping (SLAM) is a well-established
	problem in robotics and has been an active area of research over the
	past three decades \cite{Hashim2020SLAMIEEELetter,choset2000SLAM,durrant2006simultaneous,bekris2006evaluation,davison2007monoslam,zlotnik2018SLAM,hashim2021SLAMFilter}.
	The SLAM problem concerns a vehicle whose 1) pose (orientation and
	position) is unknown, traveling within 2) an unknown environment.
	This task is particularly important in GPS-denied applications, for
	instance, indoor applications, surveillance, and others. The localization
	and mapping process are performed via a set of measurements, typically,
	angular and translational velocities of the vehicle, and landmark
	measurements. It is apparent that sensor measurements are characterized
	by irregular behavior and the presence of uncertainties. Therefore,
	robust observers for SLAM are indispensable.
	
	SLAM observation is traditionally tackled using Gaussian filters or
	nonlinear observers. Gaussian filters allow to observe the vehicle's
	pose along with the surrounding landmarks. Examples of Gaussian filters
	for SLAM include the MonoSLAM approach that utilizes a single camera
	and real-time data \cite{davison2007monoslam}, FastSLAM based on
	a scalable approach \cite{montemerlo2007fastslam}, extended Kalman
	filter (EKF) \cite{huang2007convergence}, and particle filter \cite{bekris2006evaluation},
	among others. The Gaussian filters take a probabilistic approach to
	uncertainties present in measurements. It is worth mentioning that
	SLAM is an open problem and common issues are consistency \cite{dissanayake2011review},
	solution complexity \cite{cadena2016past}, and landmarks in motion.
	However, the SLAM problem is highly nonlinear and constitutes a dual
	observation process comprised of pose and environment observation.
	Pose of a vehicle is composed of: orientation (attitude) and position.
	While attitude is represented relative to the Special Orthogonal Group
	$\mathbb{SO}\left(3\right)$ \cite{hashim2018SO3Stochastic,hashim2020SO3Wiley},
	pose is described relative to the Special Euclidean Group $\mathbb{SE}\left(3\right)$
	\cite{hashim2019SE3Det,hashim2020SE3Stochastic,mohamed2019filters}.
	Gaussian filters fail to account for the high nonlinearity of the
	SLAM problem. As such, SLAM observation problem is best addressed
	using nonlinear observers. 
	
	Recent advances in the area of nonlinear observers evolved directly
	on $\mathbb{SO}\left(3\right)$ \cite{lee2012exponential,hashim2018SO3Stochastic,grip2012attitude,hashim2020SO3Wiley}
	and $\mathbb{SE}\left(3\right)$ \cite{hashim2018SE3Stochastic,zlotnik2018higher,hashim2019SE3Det,hashim2020SE3Stochastic},
	which opened the door to proposing nonlinear observers for SLAM. An
	early study that proposed using the Lie group of $\mathbb{SE}\left(3\right)$
	as the true representation of the SLAM problem was presented in \cite{strasdat2012local}.
	It was followed by two-staged observers, with nonlinear observer for
	pose estimation and Kalman filter for landmark estimation \cite{johansen2016globally}.
	The true SLAM problem is nonlinear and is modeled on the Lie group
	of $\mathbb{SLAM}_{n}\left(3\right)$. Nonlinear observers for SLAM
	on $\mathbb{SLAM}_{n}\left(3\right)$ have been proposed in \cite{Hashim2020SLAMIEEELetter,zlotnik2018SLAM,hashim2021SLAMFilter}.
	The innovative component of the observers in \cite{zlotnik2018SLAM,hashim2021SLAMFilter}
	consists in the use of constant gains which do not allow for fast
	adaptation. Accordingly, this paper proposes a nonlinear observer
	for SLAM on $\mathbb{SLAM}_{n}\left(3\right)$ that follows the structure
	of the work in \cite{Hashim2020SLAMIEEELetter,zlotnik2018SLAM} with
	the main contributions as listed below: 
	\begin{enumerate}
		\item[1)] A nonlinear observer for SLAM with fast adaptation that uses the
		available measurements of angular velocity, translational velocity,
		and landmarks.
		\item[2)] Exponential convergence of the error component is guaranteed.
		\item[3)] The closed loop error signals are guaranteed to be uniformly ultimately
		bounded.
	\end{enumerate}
	The remainder of the paper is organized as follows: Section \ref{sec:Preliminaries-and-Math}
	introduces the nomenclature, overview of $\mathbb{SO}\left(3\right)$
	and $\mathbb{SE}\left(3\right)$, and math notation. Section \ref{sec:SE3_Problem-Formulation}
	defines the SLAM problem, available sensor measurements, and the true
	motion kinematics. Section \ref{sec:SLAM_Filter} presents nonlinear
	observer for SLAM on $\mathbb{SLAM}_{n}\left(3\right)$ with fast
	adaptation. Section \ref{sec:SE3_Simulations} reveals the robustness
	of the proposed observer. Lastly, the conclusion is contained in Section
	\ref{sec:SE3_Conclusion}.
	
	\section{Preliminaries and Math Notation \label{sec:Preliminaries-and-Math}}
	
	\subsection{Nomenclature}
	
	\begin{tabular}{ll}
		$\left\{ \mathcal{I}\right\} $ & Inertial-frame\tabularnewline
		$\left\{ \mathcal{B}\right\} $  & Body-frame\tabularnewline
		$\mathbb{R}$ & Set of real numbers\tabularnewline
		$\mathbb{R}_{+}$ & Set of nonnegative real numbers\tabularnewline
		$\mathbb{R}^{n\times m}$ & Set of real numbers with dimension $n$-by-$m$\tabularnewline
		$\left\Vert y\right\Vert $ & Euclidean norm $\left\Vert y\right\Vert =\sqrt{y^{\top}y}$, $\forall y\in\mathbb{R}^{n}$\tabularnewline
		$\mathbb{SO}\left(3\right)$ & Special Orthogonal Group of order 3\tabularnewline
		$\mathbb{SE}\left(3\right)$ & Special Euclidean Group of order 3\tabularnewline
	\end{tabular}
	
	\subsection{Preliminaries}
	
	The Special Orthogonal Group $\mathbb{SO}\left(3\right)$ is described
	by 
	\[
	\mathbb{SO}\left(3\right)=\left\{ \left.R\in\mathbb{R}^{3\times3}\right|RR^{\top}=\mathbf{I}_{3}\text{, }{\rm det}\left(R\right)=+1\right\} 
	\]
	Note that $R\in\mathbb{SO}\left(3\right)$ is expressed relative to
	$\left\{ \mathcal{B}\right\} $. The Special Euclidean Group $\mathbb{SE}\left(3\right)$
	is defined as
	\[
	\mathbb{SE}\left(3\right)=\left\{ \left.\boldsymbol{T}=\left[\begin{array}{cc}
	R & P\\
	0_{1\times3} & 1
	\end{array}\right]\in\mathbb{R}^{4\times4}\right|R\in\mathbb{SO}\left(3\right),P\in\mathbb{R}^{3}\right\} 
	\]
	where $P\in\mathbb{R}^{3}$ refers to rigid-body's position. $P$
	is defined relative to $\left\{ \mathcal{I}\right\} $. $\boldsymbol{T}\in\mathbb{SE}\left(3\right)$
	is commonly known as a homogeneous transformation matrix that describes
	rigid-body's pose and is given by
	\begin{equation}
	\boldsymbol{T}=\left[\begin{array}{cc}
	R & P\\
	0_{1\times3} & 1
	\end{array}\right]\in\mathbb{SE}\left(3\right)\label{eq:T_SLAM}
	\end{equation}
	$\mathfrak{so}\left(3\right)$ is the Lie-algebra of $\mathbb{SO}\left(3\right)$
	with
	\begin{equation}
	\mathfrak{so}\left(3\right)=\left\{ \left.\left[y\right]_{\times}\in\mathbb{R}^{3\times3}\right|\left[y\right]_{\times}^{\top}=-\left[y\right]_{\times},y\in\mathbb{R}^{3}\right\} \label{eq:SLAM_so3}
	\end{equation}
	where $\left[y\right]_{\times}$ refers to a skew symmetric matrix
	such that
	\[
	\left[y\right]_{\times}=\left[\begin{array}{ccc}
	0 & -y_{3} & y_{2}\\
	y_{3} & 0 & -y_{1}\\
	-y_{2} & y_{1} & 0
	\end{array}\right]\in\mathfrak{so}\left(3\right),\hspace{1em}y=\left[\begin{array}{c}
	y_{1}\\
	y_{2}\\
	y_{3}
	\end{array}\right]
	\]
	$\mathfrak{se}\left(3\right)$ is the Lie-algebra of $\mathbb{SE}\left(3\right)$
	where {\small{}
		\[
		\mathfrak{se}\left(3\right)=\left\{ \left[U\right]_{\wedge}\in\mathbb{R}^{4\times4}\left|\exists u_{1},u_{2}\in\mathbb{R}^{3}:\left[U\right]_{\wedge}=\left[\begin{array}{cc}
		\left[u_{1}\right]_{\times} & u_{2}\\
		\underline{\mathbf{0}}_{3}^{\top} & 0
		\end{array}\right]\right.\right\} 
		\]
	}$\left[\cdot\right]_{\wedge}$ is a wedge operator that follows $\left[\cdot\right]_{\wedge}:\mathbb{R}^{6}\rightarrow\mathfrak{se}\left(3\right)$
	such that
	\begin{equation}
	\left[U\right]_{\wedge}=\left[\begin{array}{cc}
	\left[u_{1}\right]_{\times} & u_{2}\\
	\underline{\mathbf{0}}_{3}^{\top} & 0
	\end{array}\right]\in\mathfrak{se}\left(3\right),\hspace{1em}U=\left[\begin{array}{c}
	u_{1}\\
	u_{2}
	\end{array}\right]\in\mathbb{R}^{6}\label{eq:SLAM_wedge}
	\end{equation}
	Let $\left\Vert R\right\Vert _{{\rm I}}$ be a normalized Euclidean
	distance of $R\in\mathbb{SO}\left(3\right)$ where
	\begin{equation}
	\left\Vert R\right\Vert _{{\rm I}}=\frac{1}{4}{\rm Tr}\left\{ \mathbf{I}_{3}-R\right\} \in\left[0,1\right]\label{eq:SLAM_Ecul_Dist}
	\end{equation}
	For more information of attitude representation on $\mathbb{SO}\left(3\right)$
	visit \cite{hashim2018SO3Stochastic,hashim2020SO3Wiley} and pose
	representation on $\mathbb{SE}\left(3\right)$ visit \cite{hashim2019SE3Det,hashim2020SE3Stochastic}.
	
	\section{Problem Formulation\label{sec:SE3_Problem-Formulation}}
	
\begin{figure}
		\centering{}\includegraphics[scale=0.53]{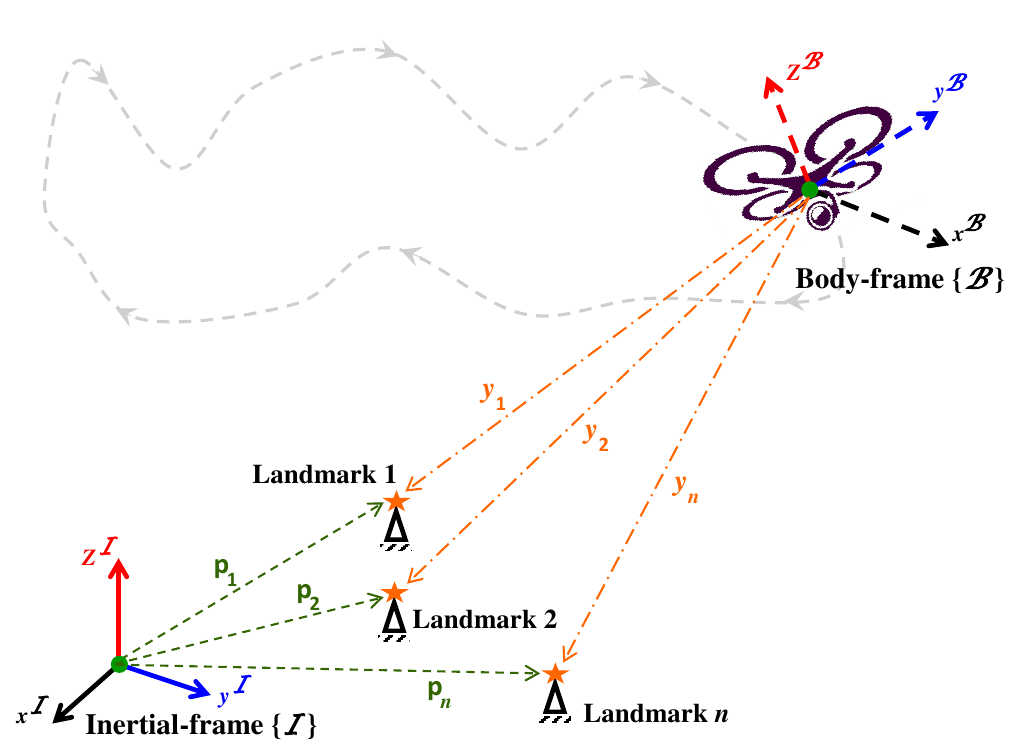}\caption{SLAM observation problem \cite{hashim2021SLAMFilter,Hashim2020SLAMIEEELetter}.}
		\label{fig:SLAM}
	\end{figure}

	SLAM estimation problem concerns simultaneous observation of the vehicle's
	pose and landmarks within the environment. Fig. \ref{fig:SLAM} illustrates
	the SLAM observation problem.
	
	Define $R\in\mathbb{SO}\left(3\right)$ as the rigid-body's attitude
	and $P\in\mathbb{R}^{3}$ as the rigid-body's translation for all
	$R\in\left\{ \mathcal{B}\right\} $ and $P\in\left\{ \mathcal{I}\right\} $.
	Assume that the map contains a family of $n$ landmarks, and let ${\rm p}_{i}$
	represent the $i$th landmark location where ${\rm p}_{i}\in\left\{ \mathcal{I}\right\} $
	for all $i=1,2,\ldots,n$. The observation problem can be solved given
	a set of measurements in the body-frame. The measurement of ${\rm p}_{i}$
	is given by
	\begin{equation}
	y_{i}=R^{\top}\left({\rm p}_{i}-P\right)+b_{i}^{y}+n_{i}^{y}\in\mathbb{R}^{3}\label{eq:SLAM_Vec_Landmark}
	\end{equation}
	where $R$, refers to the rigid-body's orientation, $P$ describes
	its translation, and ${\rm p}_{i}$ refers to the landmark's location.
	Additionally, $b_{i}^{y}$ defines unknown constant bias and $n_{i}^{y}$
	defines unknown random noise attached to the measurement with $y_{i},b_{i}^{y},n_{i}^{y}\in\left\{ \mathcal{B}\right\} $.
	
	\begin{assum}\label{Assumption:Feature}Assume three or more landmarks
		are available for measurement.\end{assum}
	
	The true motion kinematics of the rigid-body's attitude and position
	and a group of $n$-landmarks are given by \cite{hashim2021SLAMFilter,Hashim2020SLAMIEEELetter}
	\begin{align*}
	\dot{\boldsymbol{T}} & =\boldsymbol{T}\left[U\right]_{\wedge}\\
	\dot{{\rm p}}_{i} & =R{\rm v}_{i},\hspace{1em}\forall i=1,2,\ldots,n
	\end{align*}
	and in detailed form
	\begin{equation}
	\begin{cases}
	\dot{R} & =R\left[\Omega\right]_{\times}\\
	\dot{P} & =RV\\
	\dot{{\rm p}}_{i} & =R{\rm v}_{i},\hspace{1em}\forall i=1,2,\ldots,n
	\end{cases}\label{eq:SLAM_True_dot}
	\end{equation}
	where $U=$$\left[\Omega^{\top},V^{\top}\right]^{\top}$, $\Omega\in\mathbb{R}^{3}$
	defines the rigid-body's true angular velocity, $V\in\mathbb{R}^{3}$
	defines its true translational velocity, and ${\rm v}_{i}\in\mathbb{R}^{3}$
	defines the true linear velocity of the $i$th landmark. Note that
	each of $\Omega,V,{\rm v}_{i}\in\left\{ \mathcal{B}\right\} $.The
	measurements of angular and translational velocity can be described
	as
	\begin{equation}
	\begin{cases}
	\Omega_{m} & =\Omega+b_{\Omega}+n_{\Omega}\in\mathbb{R}^{3}\\
	V_{m} & =V+b_{V}+n_{V}\in\mathbb{R}^{3}
	\end{cases}\label{eq:SLAM_TVelcoity}
	\end{equation}
	where $b_{\Omega}$ defines unknown constant bias and $n_{\Omega}$
	denotes unknown random noise attached to the angular velocity, while
	$b_{V}$ defines unknown constant bias and $n_{V}$ denotes unknown
	random noise attached to the translational velocity. Note that the
	measurements of angular and translational velocities are expressed
	with respect to $\left\{ \mathcal{B}\right\} $. All landmarks are
	assumed to be fixed, thus ${\rm v}_{i}=0_{3\times1}$ $\forall i=1,2,\ldots,n$.
	
	\begin{assum}\label{Assum:Boundedness} (Uniform boundedness of $b_{\Omega}$
		and $b_{V}$) Assume that $b_{\Omega}$ and $b_{V}$ are subset of
		$\varLambda_{b}$ with $b_{\Omega},b_{V}\in\varLambda_{b}\subset\mathbb{R}^{3}$,
		where $b_{\Omega}$ and $b_{V}$ are ultimately bounded by $\Gamma$.\end{assum}
	
	\subsection{Error in Attitude, Position, and Landmark}
	
\begin{figure}
		\centering{}\includegraphics[scale=0.35]{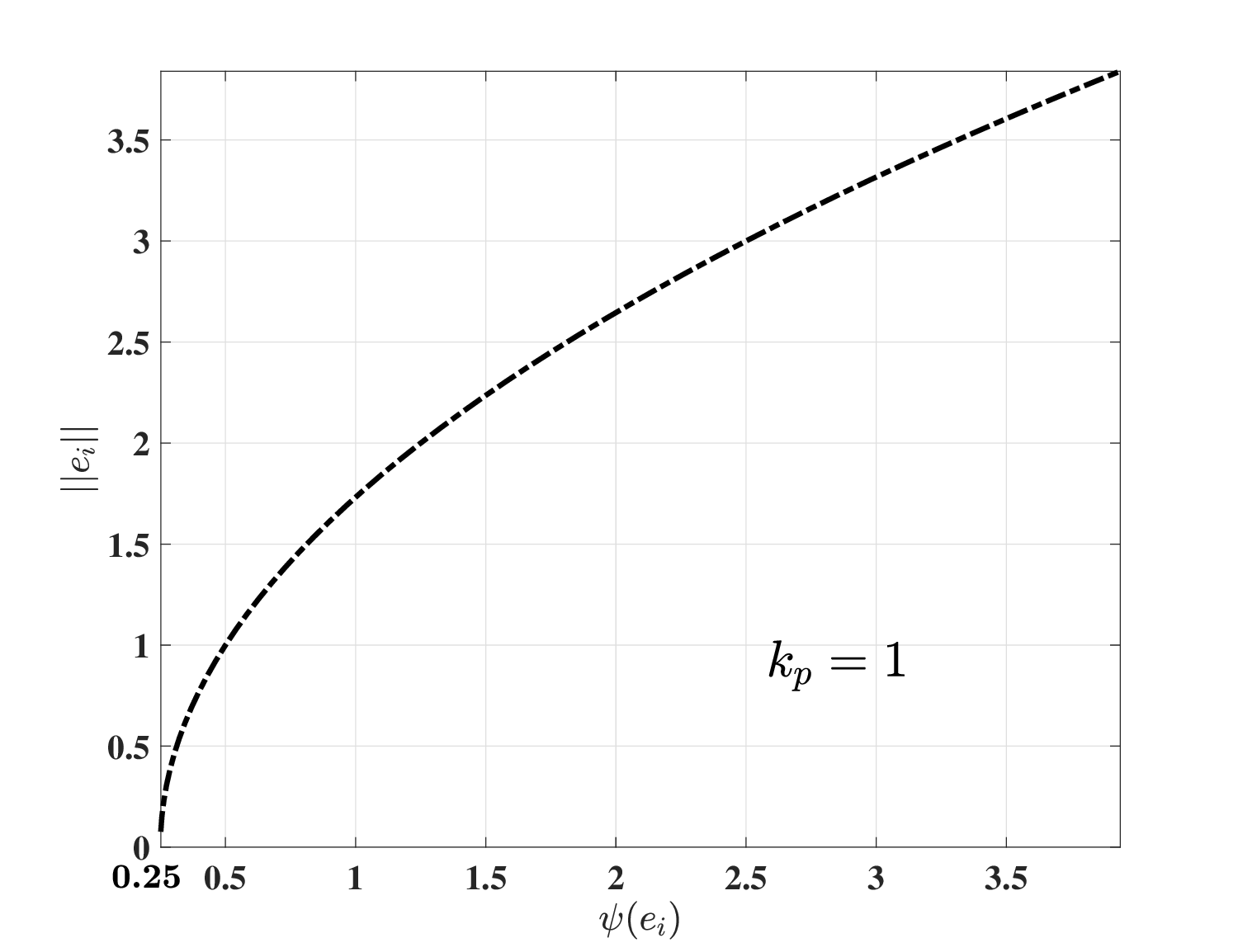}\caption{Illustrative performance of $\psi(e_{i})$: $k_{p}/4\protect\leq\psi(e_{i})<\infty$}
		\label{fig:SLAM-1}
	\end{figure}

	Consider $\hat{R}$ to be an estimate of the true orientation ($R$),
	$\hat{P}$ an estimate of the true rigid-body's position ($P$), and
	$\hat{{\rm p}}_{i}$ an estimate of the true location of the $i$th
	landmark (${\rm p}_{i}$). Consider defining the error in pose observation
	as
	\begin{align}
	\tilde{\boldsymbol{T}}=\hat{\boldsymbol{T}}\boldsymbol{T}^{-1} & =\left[\begin{array}{cc}
	\hat{R} & \hat{P}\\
	\underline{\mathbf{0}}_{3}^{\top} & 1
	\end{array}\right]\left[\begin{array}{cc}
	R^{\top} & -R^{\top}P\\
	\underline{\mathbf{0}}_{3}^{\top} & 1
	\end{array}\right]\nonumber \\
	& =\left[\begin{array}{cc}
	\tilde{R} & \tilde{P}\\
	\underline{\mathbf{0}}_{3}^{\top} & 1
	\end{array}\right]\label{eq:SLAM_T_error}
	\end{align}
	which is equivalent to
	\begin{equation}
	\begin{cases}
	\tilde{R} & =\hat{R}R^{\top}\\
	\tilde{P} & =\hat{P}-\tilde{R}P
	\end{cases}\label{eq:SLAM_T_error-1}
	\end{equation}
	Consider defining the error in the $i$th landmark observation as
	\begin{align}
	\left[\begin{array}{c}
	e_{i}\\
	0
	\end{array}\right] & =\left[\begin{array}{c}
	\hat{{\rm p}}_{i}\\
	1
	\end{array}\right]-\left[\begin{array}{cc}
	\hat{R} & \hat{P}\\
	\underline{\mathbf{0}}_{3}^{\top} & 1
	\end{array}\right]\left[\begin{array}{c}
	y_{i}\\
	1
	\end{array}\right],\hspace{1em}\forall i=1,2,\ldots,n\label{eq:SLAM_e_Final}\\
	& =\left[\begin{array}{c}
	\hat{{\rm p}}_{i}\\
	1
	\end{array}\right]-\left[\begin{array}{cc}
	\hat{R} & \hat{P}\\
	\underline{\mathbf{0}}_{3}^{\top} & 1
	\end{array}\right]\left[\begin{array}{c}
	R^{\top}\left({\rm p}_{i}-P\right)\\
	1
	\end{array}\right]\nonumber \\
	& =\left[\begin{array}{c}
	\hat{{\rm p}}_{i}\\
	1
	\end{array}\right]-\tilde{\boldsymbol{T}}\left[\begin{array}{c}
	{\rm p}_{i}\\
	1
	\end{array}\right]\nonumber \\
	& =\left[\begin{array}{c}
	\tilde{{\rm p}}_{i}-\tilde{P}\\
	0
	\end{array}\right]\nonumber 
	\end{align}
	where $\tilde{{\rm p}}_{i}=\hat{{\rm p}}_{i}-\tilde{R}{\rm p}_{i}$
	and $\tilde{P}=\hat{P}-\tilde{R}P$. Define $\hat{b}_{\Omega}$ as
	an estimate of the unknown constant bias attached to the angular velocity
	and $\hat{b}_{V}$ as an estimate of the unknown constant bias attached
	to the translational velocity. Also, consider defining the bias error
	as follows:
	\begin{equation}
	\begin{cases}
	\tilde{b}_{\Omega} & =b_{\Omega}-\hat{b}_{\Omega}\\
	\tilde{b}_{V} & =b_{V}-\hat{b}_{V}
	\end{cases}\label{eq:SLAM_b_error}
	\end{equation}
	\begin{defn}
		\label{def:Definition1}Define $x\in\mathbb{R}^{3}$ as a unit-axis
		rotating at an angle of $\theta\in\mathbb{R}$ in a 2-sphere $\mathbb{S}^{2}$.
		Angle-axis representation is one of the methods of attitude representation
		which has the map of $\mathcal{R}_{\theta}:\mathbb{R}\times\mathbb{R}^{3}\rightarrow\mathbb{SO}\left(3\right)$
		\cite{shuster1993survey,hashim2019AtiitudeSurvey}
		\begin{align*}
		\mathcal{R}_{\theta}\left(\theta,x\right) & =\mathbf{I}_{3}+\sin\left(\theta\right)\left[x\right]_{\times}+\left(1-\cos\left(\theta\right)\right)\left[x\right]_{\times}^{2}\in\mathbb{SO}\left(3\right)
		\end{align*}
		From \eqref{eq:SLAM_e_Final}, and consistently with Lemma 4 \cite{hashim2019AtiitudeSurvey},
		consider defining
		\begin{align*}
		\theta_{i} & =2\tan^{-1}(||e_{i}||)\\
		x_{i} & =\cot(\frac{\theta_{i}}{2})e_{i}
		\end{align*}
		The following mapping is obtained \cite{hashim2019AtiitudeSurvey}
		\begin{align}
		\mathcal{R}_{e(i)} & =\mathbf{I}_{3}+\sin\left(\theta_{i}\right)\left[x_{i}\right]_{\times}+\left(1-\cos\left(\theta_{i}\right)\right)\left[x_{i}\right]_{\times}^{2}\label{eq:SLAM_Compute_exp}
		\end{align}
	\end{defn}
	\begin{rem}
		\label{rem:Remark1}Recall the definition of the normalized Euclidean
		distance in \eqref{eq:SLAM_Ecul_Dist}. From \eqref{eq:SLAM_Compute_exp}
		and Definition \ref{def:Definition1}, one finds $-1\leq{\rm Tr}\{\mathcal{R}_{e(i)}\}\leq3$
		such that ${\rm Tr}\{\mathcal{R}_{e(i)}\}\rightarrow-1$ as $e_{i}\rightarrow\infty$
		and ${\rm Tr}\{\mathcal{R}_{e(i)}\}\rightarrow3$ as $e_{i}\rightarrow0$.
	\end{rem}
	\begin{defn}
		(Fast adaptation) Based on Definition \ref{def:Definition1} and Remark
		\ref{rem:Remark1}, define the following positive function $\psi:\mathbb{SO}\left(3\right)\rightarrow\mathbb{R}_{+}$
		\begin{align}
		\psi(e_{i}) & =\frac{k_{p}}{1+{\rm Tr}\{\mathcal{R}_{e(i)}\}}\label{eq:SLAM_pve_func}
		\end{align}
	\end{defn}
	The value of the function $\psi(e_{i})$ in \eqref{eq:SLAM_pve_func}
	becomes increasingly aggressive with $\psi(e_{i})\rightarrow+\infty$
	as $e_{i}\rightarrow\pm\infty$ and $\psi(e_{i})\rightarrow k_{p}/4$
	as $e_{i}\rightarrow0$, visit \cite{hashim2019AtiitudeSurvey}. The
	behavior of the proposed function in \eqref{eq:SLAM_pve_func} is
	illustrated in Fig. \ref{fig:SLAM-1}.

	\section{Nonlinear Observer Design \label{sec:SLAM_Filter}}
	
	Consider the following nonlinear observer design:
	
	\begin{equation}
	\begin{cases}
	\dot{\hat{R}} & =\hat{R}\left[\Omega_{m}-\hat{b}_{\Omega}-W_{\Omega}\right]_{\times}\\
	\dot{\hat{P}} & =\hat{R}\left(V_{m}-\hat{b}_{V}-W_{V}\right)\\
	\theta_{i} & =2\tan^{-1}(||e_{i}||),\hspace{1em}x_{i}=\cot(\frac{\theta_{i}}{2})e_{i}\\
	\mathcal{R}_{e(i)} & =\mathbf{I}_{3}+\sin\left(\theta_{i}\right)\left[x_{i}\right]_{\times}+\left(1-\cos\left(\theta_{i}\right)\right)\left[x_{i}\right]_{\times}^{2}\\
	\psi(e_{i}) & =\frac{k_{p}}{1+{\rm Tr}\{\mathcal{R}_{e(i)}\}}\\
	\dot{{\rm \hat{p}}}_{i} & =-\psi(e_{i})e_{i},\hspace{1em}\forall i=1,2,\ldots,n\\
	\dot{\hat{b}}_{\Omega} & =-\sum_{i=1}^{n}\frac{\Gamma}{\alpha_{i}}\left[y_{i}\right]_{\times}\hat{R}^{\top}e_{i}\\
	\dot{\hat{b}}_{V} & =-\sum_{i=1}^{n}\frac{\Gamma}{\alpha_{i}}\hat{R}^{\top}e_{i}\\
	W_{\Omega} & =-\sum_{i=1}^{n}\frac{k_{w}}{\alpha_{i}}\left[y_{i}\right]_{\times}\hat{R}^{\top}e_{i}\\
	W_{V} & =-\sum_{i=1}^{n}\frac{k_{w}}{\alpha_{i}}\hat{R}^{\top}e_{i}
	\end{cases}\label{eq:Filter1}
	\end{equation}
	where $k_{w}\in\mathbb{R}$, $k_{p}\in\mathbb{R}$, $\Gamma\in\mathbb{R}^{3\times3}$,
	and $\alpha_{i}\in\mathbb{R}$ are positive constants, $W_{\Omega}$
	and $\ensuremath{W_{V}}$ are correction factors, and $\text{\ensuremath{\hat{b}_{\Omega}}}$
	and $\hat{b}_{V}$ are the estimates of $b_{\Omega}$ and $b_{V}$,
	respectively.
	\begin{thm}
		Consider the true motion kinematics in \eqref{eq:SLAM_True_dot},
		landmark measurements ($y_{i}=R^{\top}\left({\rm p}_{i}-P\right)$)
		for all $i=1,2,\ldots,n$, and angular velocity measurement $\Omega_{m}=\Omega+b_{\Omega}$,
		and translational velocity measurement $V_{m}=V+b_{V}$ as in \eqref{eq:SLAM_TVelcoity}.
		Assume that Assumption \ref{Assumption:Feature} is met. Consider
		the observer design to be as in \eqref{eq:Filter1}. Define the set
		\begin{align}
		\mathcal{S}= & \left\{ \left(e_{1},e_{2},\ldots,e_{n}\right)\in\mathbb{R}^{3}\times\mathbb{R}^{3}\times\cdots\times\mathbb{R}^{3}\right|\nonumber \\
		& \hspace{10em}\left.e_{i}=\underline{\mathbf{0}}_{3}\forall i=1,2,\ldots n\right\} \label{eq:SLAM_Set1}
		\end{align}
		Then
	\end{thm}
	\begin{itemize}
		\item[1)] $e_{i}$ in \eqref{eq:SLAM_e_Final} exponentially approaches $\mathcal{S}$,
		and
		\item[2)]  the error in attitude and position $\tilde{R}\rightarrow R_{c}$
		and $\tilde{P}\rightarrow P_{c}$ as $t\rightarrow\infty$ where $R_{c}\in\mathbb{SO}\left(3\right)$
		refers to a constant matrix and $P_{c}\in\mathbb{R}^{3}$ refers to
		a constant vector.
	\end{itemize}
	\begin{proof}From \eqref{eq:SLAM_T_error}, one has
		\begin{align}
		\dot{\tilde{\boldsymbol{T}}} & =\dot{\hat{\boldsymbol{T}}}\boldsymbol{T}^{-1}+\hat{\boldsymbol{T}}\dot{\boldsymbol{T}}^{-1}\nonumber \\
		& =\hat{\boldsymbol{T}}\left[\tilde{b}_{U}-W_{U}\right]_{\wedge}\hat{\boldsymbol{T}}^{-1}\tilde{\boldsymbol{T}}\label{eq:SLAM_T_error_dot}
		\end{align}
		where $\boldsymbol{\dot{T}}^{-1}=-\boldsymbol{T}^{-1}\boldsymbol{\dot{T}}\boldsymbol{T}^{-1}$.
		Thereby, the error dynamics of \eqref{eq:SLAM_e_Final} are equivalent
		to
		\begin{align}
		\left[\begin{array}{c}
		\dot{e}_{i}\\
		0
		\end{array}\right] & =\left[\begin{array}{c}
		\dot{{\rm \hat{p}}}_{i}\\
		0
		\end{array}\right]-\dot{\tilde{\boldsymbol{T}}}\left[\begin{array}{c}
		{\rm p}_{i}\\
		1
		\end{array}\right]-\tilde{\boldsymbol{T}}\,\dot{\overline{{\rm p}}}_{i}\nonumber \\
		& =\left[\begin{array}{c}
		\dot{{\rm \hat{p}}}_{i}\\
		0
		\end{array}\right]-\hat{\boldsymbol{T}}\left[\tilde{b}_{U}-W_{U}\right]_{\wedge}\hat{\boldsymbol{T}}^{-1}\tilde{\boldsymbol{T}}\left[\begin{array}{c}
		{\rm p}_{i}\\
		1
		\end{array}\right]\label{eq:SLAM_e_dot}
		\end{align}
		From \eqref{eq:SLAM_T_error_dot}, one finds
		\begin{align}
		\hat{\boldsymbol{T}}\left[\tilde{b}_{U}\right]_{\wedge}\hat{\boldsymbol{T}}^{-1} & =\left[\begin{array}{c}
		\hat{R}\tilde{b}_{\Omega}\\
		\hat{R}\tilde{b}_{V}+\left[\hat{P}\right]_{\times}\hat{R}\tilde{b}_{\Omega}
		\end{array}\right]_{\wedge}\in\mathfrak{se}\left(3\right)\label{eq:SLAM_Adj_Property4}
		\end{align}
		where for $x\in\mathbb{R}^{3}$ and $R\in\mathbb{SO}(3)$, $\left[Rx\right]_{\times}=R\left[x\right]_{\times}R^{\top}$.
		Accordingly, the result in \eqref{eq:SLAM_Adj_Property4} is equivalent
		to
		\begin{equation}
		\hat{\boldsymbol{T}}\left[\tilde{b}_{U}\right]_{\wedge}\hat{\boldsymbol{T}}^{-1}=\left[\left[\begin{array}{cc}
		\hat{R} & 0_{3\times3}\\
		\left[\hat{P}\right]_{\times}\hat{R} & \hat{R}
		\end{array}\right]\left[\begin{array}{c}
		\tilde{b}_{\Omega}\\
		\tilde{b}_{V}
		\end{array}\right]\right]_{\wedge}\label{eq:SLAM_Adj_Property5}
		\end{equation}
		which shows that
		\begin{align}
		\hat{\boldsymbol{T}}\left[\tilde{b}_{U}\right]_{\wedge}\hat{\boldsymbol{T}}^{-1}\tilde{\boldsymbol{T}}\,\overline{{\rm p}}_{i} & =\left[\begin{array}{cc}
		-\hat{R}\left[y_{i}\right]_{\times} & \hat{R}\\
		\underline{\mathbf{0}}_{3}^{\top} & \underline{\mathbf{0}}_{3}^{\top}
		\end{array}\right]\left[\begin{array}{c}
		\tilde{b}_{\Omega}\\
		\tilde{b}_{V}
		\end{array}\right]\label{eq:SLAM_Adj_Property6}
		\end{align}
		Therefore, it can be concluded that the error dynamics are
		\begin{align}
		\dot{e}_{i} & =\dot{\hat{{\rm p}}}_{i}-\left[\begin{array}{cc}
		-\hat{R}\left[y_{i}\right]_{\times} & \hat{R}\end{array}\right]\left[\begin{array}{c}
		\tilde{b}_{\Omega}-W_{\Omega}\\
		\tilde{b}_{V}-W_{V}
		\end{array}\right]\label{eq:SLAM_e_dot_Final}
		\end{align}
		Consider the candidate Lyapunov function $\boldsymbol{{\rm V}}=\boldsymbol{{\rm V}}\left(e_{1},\ldots,e_{n},\tilde{b}_{\Omega},\tilde{b}_{V}\right)$
		defined as follows:
		\begin{equation}
		\boldsymbol{{\rm V}}=\sum_{i=1}^{n}\frac{1}{2\alpha_{i}}e_{i}^{\top}e_{i}+\frac{1}{2}\tilde{b}_{\Omega}^{\top}\Gamma^{-1}\tilde{b}_{\Omega}+\frac{1}{2}\tilde{b}_{V}^{\top}\Gamma^{-1}\tilde{b}_{V}\label{eq:SLAM_Lyap1}
		\end{equation}
		The time derivative of \eqref{eq:SLAM_Lyap1} becomes
		\begin{align}
		\dot{\boldsymbol{{\rm V}}}= & \sum_{i=1}^{n}\frac{1}{\alpha_{i}}e_{i}^{\top}\dot{e}_{i}-\tilde{b}_{\Omega}^{\top}\Gamma^{-1}\dot{\hat{b}}_{\Omega}-\tilde{b}_{V}^{\top}\Gamma^{-1}\dot{\hat{b}}_{V}\nonumber \\
		= & -\sum_{i=1}^{n}\frac{1}{\alpha_{i}}e_{i}^{\top}\left[\begin{array}{cc}
		-\hat{R}\left[y_{i}\right]_{\times} & \hat{R}\end{array}\right]\left[\begin{array}{c}
		\tilde{b}_{\Omega}-W_{\Omega}\\
		\tilde{b}_{V}-W_{V}
		\end{array}\right]\nonumber \\
		& +\sum_{i=1}^{n}\frac{1}{\alpha_{i}}e_{i}^{\top}\dot{\hat{{\rm p}}}_{i}-\tilde{b}_{\Omega}^{\top}\Gamma^{-1}\dot{\hat{b}}_{\Omega}-\tilde{b}_{V}^{\top}\Gamma^{-1}\dot{\hat{b}}_{V}\nonumber \\
		= & \sum_{i=1}^{n}\frac{1}{\alpha_{i}}e_{i}^{\top}\hat{R}\left[y_{i}\right]_{\times}(\tilde{b}_{\Omega}-W_{\Omega})\nonumber \\
		& -\sum_{i=1}^{n}\frac{1}{\alpha_{i}}e_{i}^{\top}\hat{R}(\tilde{b}_{V}-W_{V})\nonumber \\
		& +\sum_{i=1}^{n}\frac{1}{\alpha_{i}}e_{i}^{\top}\dot{\hat{{\rm p}}}_{i}-\tilde{b}_{\Omega}^{\top}\Gamma^{-1}\dot{\hat{b}}_{\Omega}-\tilde{b}_{V}^{\top}\Gamma^{-1}\dot{\hat{b}}_{V}\label{eq:SLAM_Lyap1_dot}
		\end{align}
		Replacing $W_{\Omega}$, $W_{V}$, $\dot{\hat{b}}_{\Omega}$, $\dot{\hat{b}}_{V}$,
		and $\dot{\hat{{\rm p}}}_{i}$ with their definitions in \eqref{eq:Filter1}
		leads to
		\begin{align}
		\dot{\boldsymbol{{\rm V}}}\leq & -\sum_{i=1}^{n}\frac{\psi(e_{i})}{\alpha_{i}}\left\Vert e_{i}\right\Vert ^{2}-k_{w}\left\Vert \sum_{i=1}^{n}\frac{e_{i}}{\alpha_{i}}\right\Vert ^{2}\label{eq:SLAM_Lyap1_dot_Final}
		\end{align}
		From \eqref{eq:SLAM_Lyap1_dot_Final}, $\boldsymbol{{\rm V}}$ is
		negative for all $e_{i}\neq0$ and $\boldsymbol{{\rm V}}$ is equal
		to zero at $e_{i}=0_{3\times1}$. Thus, the inequality in \eqref{eq:SLAM_Lyap1_dot_Final}
		shows that $e_{i}$ is regulated exponentially to the set $\mathcal{S}$
		in \eqref{eq:SLAM_Set1}. In view of Barbalat Lemma, $\dot{\boldsymbol{{\rm V}}}$
		is negative, continuous, and converges to zero indicating that $\tilde{b}_{\Omega}$
		and $\tilde{b}_{V}$ are bounded. As such, $\tilde{R}\rightarrow R_{c}$
		and $\tilde{P}\rightarrow P_{c}$ as $t\rightarrow\infty$ completing
		the proof.\end{proof}
	
	The discrete implementation of the observer in \eqref{eq:Filter1}
	is given by{\small{}
		\begin{equation}
		\begin{cases}
		\hat{\boldsymbol{T}}[k+1] & =\hat{\boldsymbol{T}}[k]\exp\left(\left[\begin{array}{c}
		\Omega_{m}[k]-\hat{b}_{\Omega}[k]-W_{\Omega}[k]\\
		V_{m}[k]-\hat{b}_{V}[k]-W_{V}[k]
		\end{array}\right]_{\wedge}\Delta t\right)\\
		\theta_{i} & =2\tan^{-1}(||e_{i}[k]||),\hspace{1em}x_{i}=\cot(\frac{\theta_{i}}{2})e_{i}[k]\\
		\mathcal{R}_{e(i)} & =\mathbf{I}_{3}+\sin\left(\theta_{i}\right)\left[x_{i}\right]_{\times}+\left(1-\cos\left(\theta_{i}\right)\right)\left[x_{i}\right]_{\times}^{2}\\
		\psi(e_{i}) & =\frac{k_{p}}{1+{\rm Tr}\{\mathcal{R}_{e(i)}\}}\\
		{\rm \hat{p}}_{i}[k+1] & ={\rm \hat{p}}_{i}[k]-\Delta t\psi(e_{i})e_{i}[k],\hspace{1em}\forall i=1,2,\ldots,n\\
		\hat{b}_{\Omega}[k+1] & =\hat{b}_{\Omega}[k]-\Delta t\sum_{i=1}^{n}\frac{\Gamma}{\alpha_{i}}\left[y_{i}[k]\right]_{\times}\hat{R}^{\top}[k]e_{i}[k]\\
		\hat{b}_{V}[k+1] & =\hat{b}_{V}[k]-\sum_{i=1}^{n}\frac{\Gamma}{\alpha_{i}}\hat{R}^{\top}[k]e_{i}[k]\\
		W_{\Omega} & =-\sum_{i=1}^{n}\frac{k_{w}}{\alpha_{i}}\left[y_{i}[k]\right]_{\times}\hat{R}^{\top}[k]e_{i}[k]\\
		W_{V} & =-\sum_{i=1}^{n}\frac{k_{w}}{\alpha_{i}}\hat{R}^{\top}[k]e_{i}[k]
		\end{cases}\label{eq:Filter1-1}
		\end{equation}
	}{\small\par}
	
	\section{Simulation Results \label{sec:SE3_Simulations}}
	
	This section reveals the robustness of the proposed nonlinear observer
	with fast adaptation for SLAM on the Lie group of $\mathbb{SLAM}_{n}\left(3\right)$.
	Consider the following set of data, initialization parameters, and
	measurement bias: 
	\[
	\begin{cases}
	\Omega & =[0,0,0.3]^{\top}({\rm rad/sec})\\
	V & =[2.5,0,0]^{\top}({\rm m/sec})\\
	R\left(0\right) & =\mathbf{I}_{3}\\
	P\left(0\right) & =[0,0,6]^{\top}\\
	{\rm p}_{1} & =[7,7,0]^{\top}\\
	{\rm p}_{2} & =[-7,7,0]^{\top}\\
	{\rm p}_{3} & =[7,-7,0]^{\top}\\
	{\rm p}_{4} & =[-7,-7,0]^{\top}\\
	b_{\Omega} & =[0.09,-0.15,-0.1]^{\top}({\rm rad/sec})\\
	b_{V} & =[0.09,0.06,-0.07]^{\top}({\rm m/sec})
	\end{cases}
	\]
	Consider the initial estimates of attitude, position, and landmark
	locations to be
	\begin{align*}
	\hat{R}\left(0\right) & =\mathbf{I}_{3},\hspace{1em}\hat{P}\left(0\right)=0_{3\times1}\\
	\hat{{\rm p}}_{1}\left(0\right) & =\hat{{\rm p}}_{2}\left(0\right)=\hat{{\rm p}}_{3}\left(0\right)=\hat{{\rm p}}_{4}\left(0\right)=0_{3\times1}
	\end{align*}
	Consider selecting the design parameters as follows: $\alpha_{i}=0.1$,
	$\Gamma=30\mathbf{I}_{3}$, $k_{p}=1$, and $k_{w}=2$, while the
	initial estimates of the biases are $\hat{b}_{\Omega}\left(0\right)=\hat{b}_{V}\left(0\right)=0_{3\times1}$
	for all $i=1,2,3,4$.
	
\begin{figure}[h]
		\centering{}\includegraphics[scale=0.34]{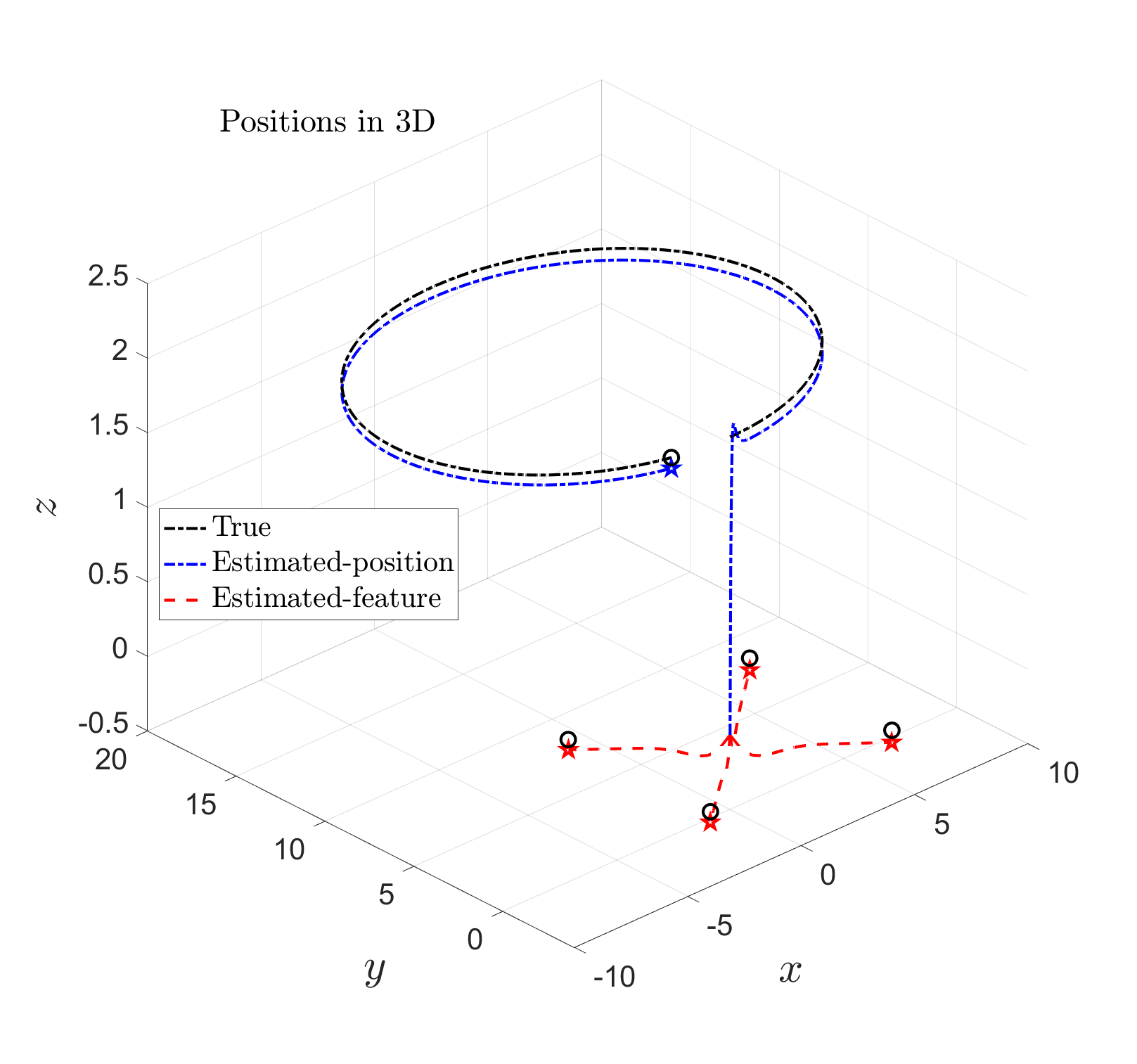}\caption{Output performance of fast adaptation nonlinear observer for SLAM
			vs True}
		\label{fig:SLAM_3d}
	\end{figure}

	Fig. \ref{fig:SLAM_3d} illustrates the output performance of the
	proposed observer against the true trajectory. The true performance
	is depicted in black center-line with final destinations depicted
	as black circles. The estimated performance is shown in red dash-line
	and blue center-line, while the estimated positions of the final destinations
	are depicted as red and blue stars $\star$. Fig. \ref{fig:SLAM_3d}
	reveals strong estimation capabilities of the proposed observer in
	localizing the unknown pose of the vehicle as well as mapping the
	unknown environment.
	
\begin{figure}[h]
		\centering{}\includegraphics[scale=0.37]{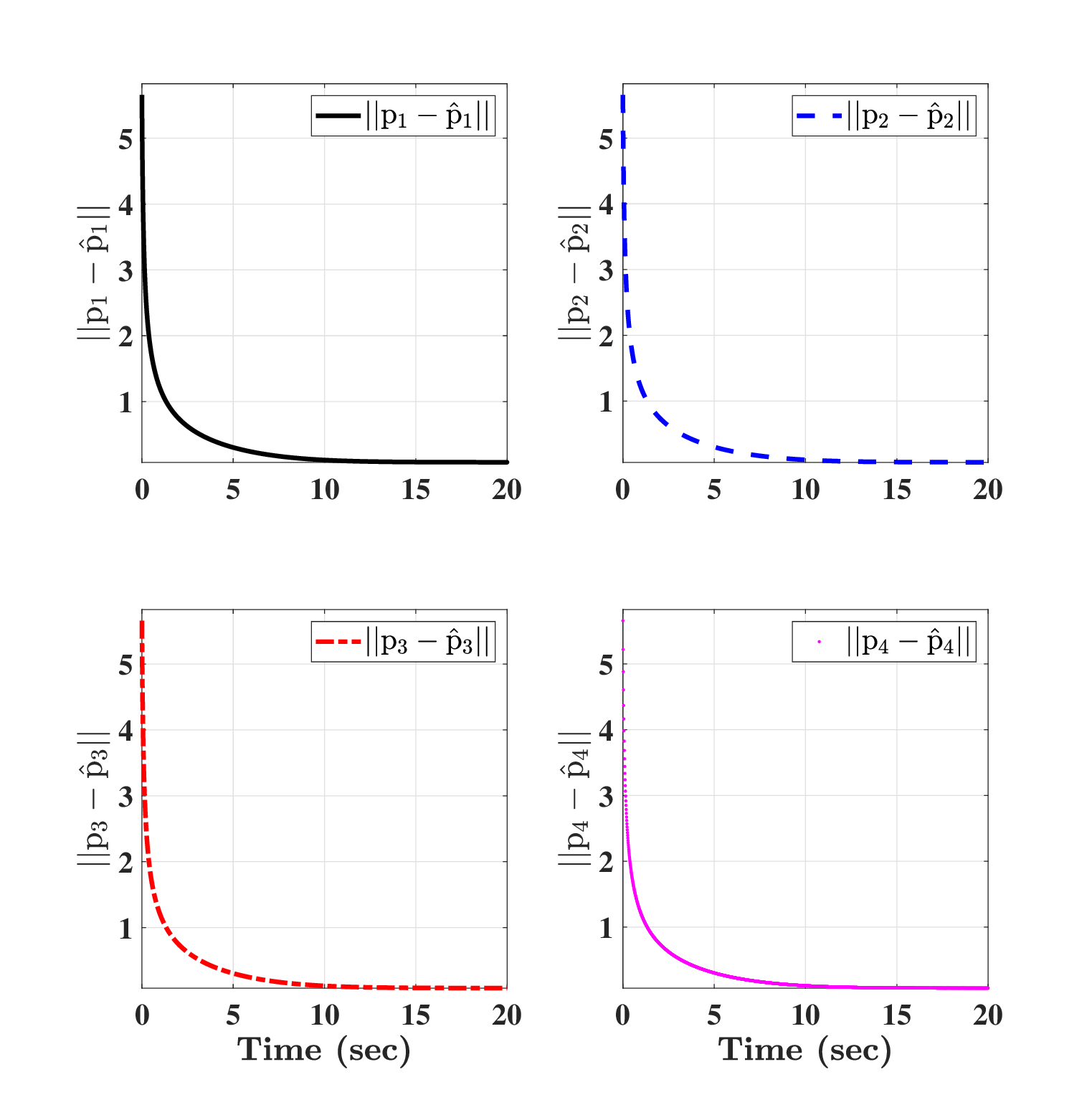}\caption{Error trajectories of $||{\rm p}_{i}-{\rm \hat{p}}_{i}||$.}
		\label{fig:SLAM_error_p}
	\end{figure}

\begin{figure}[h]
		\centering{}\includegraphics[scale=0.37]{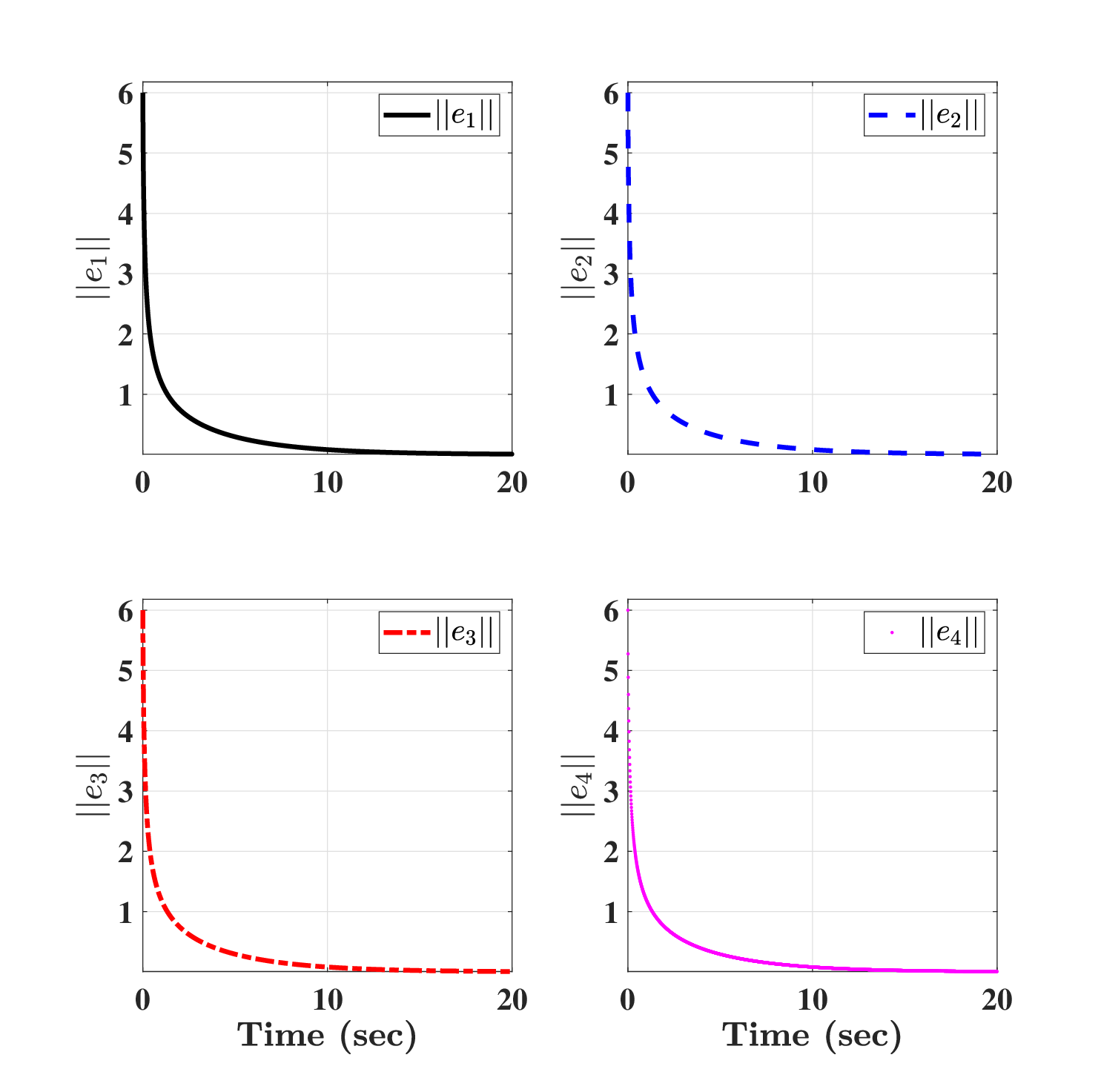}\caption{Evolution of error trajectories of $e_{i}$.}
		\label{fig:SLAM_error_Lyap}
	\end{figure}

	Fig. \ref{fig:SLAM_error_p} reveals asymptotic and fast convergence
	of $e_{i}$ to the origin from large error in initialization. Likewise,
	Fig. \ref{fig:SLAM_error_Lyap} demonstrates fast convergence of $||{\rm p}_{i}-{\rm \hat{p}}_{i}||$
	from large error in initialization to the close neighborhood of the
	origin.
	
	\section{Conclusion \label{sec:SE3_Conclusion}}
	
	A nonlinear observer for Simultaneous Localization and Mapping (SLAM)
	modeled on the Lie group of $\mathbb{SLAM}_{n}\left(3\right)$ is
	proposed. The observer follows the structure of the true SLAM problem.
	The proposed observer compensates for the unknown bias attached to
	angular and translational velocities. The proposed observer can be
	easily implemented on a vehicle given the availability of velocity
	and landmark measurements. Numerical results revealed the observer's
	ability to concurrently map the unknown environment and obtain the
	vehicle's pose.
	
	\section*{Acknowledgment}
	
	The authors would like to thank \textbf{Maria Shaposhnikova} for proofreading
	the article.

	\bibliographystyle{IEEEtran}
	\bibliography{bib_SLAM}

\end{document}